\documentclass[]{webofc}
\usepackage[varg]{txfonts}   

\usepackage{mathrsfs}
\usepackage{amssymb}
\usepackage{amsmath}
\usepackage{graphicx}
\usepackage{hyperref}
\usepackage{amsfonts,amsbsy}
\usepackage{nicefrac}
\usepackage{xcolor}

\usepackage{comment}
\usepackage{bbm}
\usepackage{slashed}

\def\bea{\begin{eqnarray}}
\def\eea{\end{eqnarray}}

\def\be{\begin{equation}}
\def\ee{\end{equation}}

\newcommand{\SU}{\mrm{SU}}

\newcommand{\mbf}{\mathbf}

\newcommand{\tpert}{t}
\newcommand{\mrm}{\mathrm}

\newcommand{\HTL}{\mrm{HTL}}

\newcommand{\Q}{Q}
\newcommand{\wplas}{\omega_{\mrm{pl}}}

\newcommand{\fig}{Fig.~}
\newcommand{\figs}{Figs.~}
\newcommand{\eq}{Eq.~}
\newcommand{\se}{Sec.~}

\newcommand{\re}{Ref.~}
\newcommand{\res}{Refs.~}

\newcommand{\non}{\nonumber \\}

\newcommand{\ud}{\mathrm{d}}

\newcommand{\pToFigs}{.}

\begin{document}
\title{Fermion and gluon spectral functions far from equilibrium}
%
%

\author{
\firstname{Kirill} \lastname{Boguslavski}\inst{1}\fnsep\thanks{Speaker,\quad \email{kirill.boguslavski@tuwien.ac.at}. \quad 
\newline K.B.\ would like to acknowledge support by the Austrian Science Fund (FWF) under project P 34455-N. 
\newline The authors are grateful to A.~Kurkela, M.~Mace and J.~Peuron for collaboration and discussions. 
The authors wish to acknowledge the National Energy Research Scientific Computing Center, a U.S. Department of Energy Office of Science User Facility supported under Contract No. DE-AC02-05CH11231, the CSC - IT Center for Science, Finland, and the Vienna Scientific Cluster (VSC, Austria) under the project number 71444 for computational resources. 
} \and
\firstname{Tuomas} \lastname{Lappi}\inst{2,3}\fnsep\thanks{
T.L.\ is supported by the Academy of Finland, project 321840 and under the European Union’s Horizon 2020 research and innovation programme by the European Research Council (ERC, grant agreement No. ERC-2015-CoG-681707) and by the STRONG-2020 project (grant agreement No 824093). The content of this article does not reflect the official opinion of the European Union and responsibility for the information and views expressed therein lies entirely with the authors.
}
\and
\firstname{S\"oren} \lastname{Schlichting}\inst{4}\fnsep\thanks{
S.S.\ is supported by the Deutsche Forschungsgemeinschaft (DFG, German Research Foundation) through the CRC-TR 211 ``Strong-interaction matter under extreme conditions'' Project number 315477589.
}
}

\institute{Institute  for  Theoretical  Physics,  Technische  Universität  Wien,  1040  Vienna,  Austria
\and
Department of Physics, P.O.~Box 35, 40014 University of Jyv\"{a}skyl\"{a}, Finland
\and
Helsinki Institute of Physics, P.O.~Box 64, 00014 University of Helsinki, Finland
\and
Fakultät für Physik, Universität Bielefeld, D-33615 Bielefeld, Germany
          }

\abstract{%
  Motivated by the quark-gluon plasma, we develop a simulation method to obtain the spectral function of (Wilson) fermions non-perturbatively in a non-Abelian gauge theory with large gluon occupation numbers \cite{Boguslavski:2021kdd}. We apply our method to a non-Abelian plasma close to its non-thermal fixed point, i.e., in a far-from-equilibrium self-similar regime, and find mostly very good agreement with perturbative hard loop (HTL) calculations. For the first time, we extract the full momentum dependence of the damping rate of fermionic collective excitations and compare our results to recent non-perturbative extractions of gluonic spectral functions in two and three spatial dimensions \cite{Boguslavski:2021buh,Boguslavski:2018beu}. 
}

\maketitle
%

\section{Introduction}
\label{intro}

Nonperturbatively strong bosonic fields can be encountered in a variety of systems, including heavy-ion collisions at early times \cite{Schlichting:2019abc,Berges:2020fwq}, cosmological reheating \cite{Allahverdi:2010xz}, or the infrared sector of gluonic and scalar fields in thermal equilibrium. Understanding their interactions with fermions is important for phenomenological applications including jets (energy loss, jet quenching) or electromagnetic observables, to mention a few in heavy-ion collisions. 

For a microscopic description of 
the dynamics, the calculation of spectral functions $\rho(\omega,p)$ is very useful, since they encode medium interactions and the spectrum of collective excitations. 

In this talk we will present a new method to compute the spectral function of Wilson fermions non-perturbatively in a weakly-coupled non-Abelian gauge theory with large gluon occupation numbers $f \gg 1$ (\se\ref{sec:method}). In this far-from-equilibrium situation, we can use classical-statistical lattice simulations to describe the underlying dynamics \cite{Berges:2004yj,Berges:2013lsa}. Coupling them to the Dirac equation allows us to extract the fermion spectral function nonperturbatively. Our results are then discussed and compared to perturbative hard-loop (HTL) computations (\se\ref{sec:fermion_rho}) and to gluonic spectral functions in different dimensions (\se\ref{sec:gluon_rhos}, for the comparison see \se\ref{sec:conclusion}). We conclude in \se\ref{sec:conclusion}. More details on our method and results for the fermion spectral function can be found in \re\cite{Boguslavski:2021kdd}. Our discussion of the gluon spectral function results is based on \res\cite{Boguslavski:2021buh,Boguslavski:2018beu}.


\section{Method \& setup}
\label{sec:method}


\subsection{Theory}

We consider $\SU(N_c)$ gauge theory on a cubic lattice of size $N_s^3$ and lattice spacing $a_s$. The Yang-Mills and Dirac Hamiltonians read in $A_0=0$ gauge
\begin{align}
 H_{\text YM} =& \frac{1}{g^2 a_s}\sum_{\mbf x,i}  \text{Tr}[E_{i}(t',\mbf x)^2] + \frac{1}{2}\sum_{j} \text{ReTr}[1-U_{ij}(t',\mbf x)] \non
 \hat{H}_W =& \frac{1}{2} \sum_{\mbf x} \left[ \hat{\psi}^\dagger(t',\mbf x), \gamma^0 \left( -i\slashed{D}_s[U] + m \right) \hat{\psi}(t',\mbf x) \right],
\end{align}
with the link fields $U_j \approx \exp\left( ig\, a_s A_j \right)$ replacing gauge fields, plaquettes $U_{ij}$ computed from link fields, and $-i\slashed{D}_s[U] \hat{\psi}$ being the tree-level improved Wilson Dirac operator \cite{Mace:2016shq,Mace:2019cqo}. We neglect the backreaction of fermions on the evolution of gauge fields since it is subleading when gluonic occupation numbers are large, as in our case. 
We will consider $N_c=2$ in the following to reduce the computational complexity of the calculations.

The fermionic fields are formulated in terms of a mode expansion \cite{Aarts:1998td}
\be
 \label{eq:mode_exp}
 \hat{\psi}(t',\mbf x) = \frac{1}{\sqrt{V}} \sum_{\lambda,\mbf p} \hat{b}_{\lambda,\mbf p}(\tpert)\,  \phi_{\lambda,\mbf p}^{u}(t',\mbf x) +  \hat{d}^{\dagger}_{\lambda,\mbf p}(\tpert)\,  \phi_{\lambda,\mbf p}^{v}(t',\mbf x),
\ee
where $\lambda=1,\cdots,2N_c$ labels spin and color indices. At the reference time $t' = \tpert$, where $\tpert$ will be the time when we extract the spectral function, the creation and annihilation operators $\hat{b}$, $\hat{d}$ satisfy the usual anti-commutation relations at $t$ and the wave functions are set to plane waves 
\be
 \label{eq:plane_waves}
 \left.\phi_{\lambda,\mbf p}^{u}(t',\mbf x)\right|_{t'{=}\tpert} = u_{\lambda}(\mbf p)e^{+ i \mbf p \cdot \mbf x}\,, \quad
 \left.\phi_{\lambda,\mbf p}^{v}(t',\mbf x)\right|_{t'{=}\tpert} = v_{\lambda}(\mbf p)e^{-i \mbf p \cdot \mbf x}\,.
\ee


\subsection{Fermion spectral function}

The spectral function of fermions is defined as
\be
 \label{eq:rho_ferm}
 \rho^{\alpha \beta}(x,y) = \left\langle\left\lbrace \hat{\psi}^{\alpha}(t',\mbf x), \hat{\bar{\psi}}^{\beta}(\tpert,\mbf y) \right\rbrace\right\rangle,
\ee
where we have written the Dirac indices $\alpha$, $\beta$ explicitly. In our classical-statistical framework, the spectral function can be computed efficiently in momentum space. Plugging in the mode expansion \eqref{eq:mode_exp} into \eqref{eq:rho_ferm}, Fourier transforming with respect to $\mbf x - \mbf y$, and using \eqref{eq:plane_waves} for time $\tpert$, one arrives at
\begin{align}
 \label{eq:fermion_rho_calc}
 & \rho^{\alpha \beta}(t',\tpert,\mbf p) \\
 &= \frac{1}{V}\sum_{\lambda,\mbf q} \left\langle \tilde{\phi}^{u,\alpha}_{\lambda,\mbf q}(t',\mbf p) \left( \tilde{\phi}^{u,\gamma}_{\lambda,\mbf q}(\tpert,\mbf p) \right)^{*} + \tilde{\phi}^{v,\alpha}_{\lambda,\mbf q}(t',\mbf p) \left( \tilde{\phi}^{v,\gamma}_{\lambda,\mbf q}(\tpert,\mbf p) \right)^{*} \right\rangle \gamma_{0}^{\gamma\beta} \non
 &= \frac{1}{V}\sum_{\lambda} \left\langle \tilde{\phi}^{u,\alpha}_{\lambda,\mbf p}(t',\mbf p)\, u_{\lambda}^{\dagger,\gamma}(\mbf p) + \tilde{\phi}^{v,\alpha}_{\lambda,-\mbf p}(t',\mbf p)\, v_{\lambda}^{\dagger,\gamma}(-\mbf p) \right\rangle \gamma_{0}^{\gamma\beta}\,, \nonumber
\end{align}
where $\langle . \rangle$ denotes a classical-statistical average over gluonic configurations.
The initialization and the specific form of the initial conditions in \eq\eqref{eq:plane_waves}, where $\phi_{\lambda,\mbf q}^{u/v}(\tpert,\mbf x)$ is given by a plane wave, 
allows to simplify the expression in the last line above.


\subsection{Simulation algorithm}
\label{sec:simul_alg_fermions}

The simulation algorithm can be summarized as follows:

\begin{enumerate}
   
   \item Set {\em initial conditions} for gluons at $t' = 0$, generating a configuration with
   $\langle E_T^*(t'{=}0,\mbf p) E_T(t'{=}0,\mbf q) \rangle \propto p\,f(t'{=}0,p) (2\pi)^3\delta(\mbf p - \mbf q)$.
   
   \item Solve classical equations of motion (EOMs) for {\em gauge fields} for $0 \leq t' \leq \tpert$, set Coulomb-type gauge $\left. \partial^j A_j\right|_{\tpert} = 0$ at $t' = \tpert$, i.e., at the time when the spectral function is measured.
   
   \item For each momentum mode $\mbf p$ initialize $\phi_{\lambda,\mbf p}^{u/v}$ at $t' = \tpert$ using \eq\eqref{eq:plane_waves}.
   
  \item {\em Evolve gauge and fermionic fields} for $t' > t$ using a leap-frog scheme to solve the classical EOMs for gauge and the Dirac equation for fermionic fields.
  
  \item Calculate the fermionic spectral function $\rho(t',\tpert,\mbf p)$ for each momentum mode $\mbf p$ according to \eq\eqref{eq:fermion_rho_calc}.
  
\end{enumerate}

\section{Nonperturbatively computed fermion spectral functions}
\label{sec:fermion_rho}


\subsection{Non-equilibrium state: self-similar turbulent attractor}

We now consider spectral functions in a state of high gluonic occupancy $f_g(t{=}0, p \lesssim \Q) \sim \frac{1}{g^2} \gg 1$. 
Such a system approaches a self-similar attractor with
\begin{align}
 \label{eq:self-sim}
 f(t,p) = (\Q t)^{\alpha} f_s\left((\Q t)^{\beta} p \right),
\end{align}
after some short transient time. The scaling exponents and scaling function then become insensitive to details of initial conditions and their universal values in Minkowski space-time are given by $\beta = -1/7$ and $\alpha = 4\beta$ for a 3+1 dimensional \cite{Kurkela:2012hp,Schlichting:2012es,Berges:2013fga} and by $\beta = -1/5$ and $\alpha = 3\beta$ for a 2+1D plasma \cite{Boguslavski:2019fsb}. Since we know the time-evolution of characteristic momentum scales at those attractor states, like the hard scale $\Lambda \sim Q(Q t)^{-\beta}$ and the soft mass scale $m \sim Q(Q t)^{\beta}$, these states are useful assets to study spectral functions in more detail. 


\begin{figure}[t!]
 \centering
 \includegraphics[width=0.47\textwidth]{\pToFigs/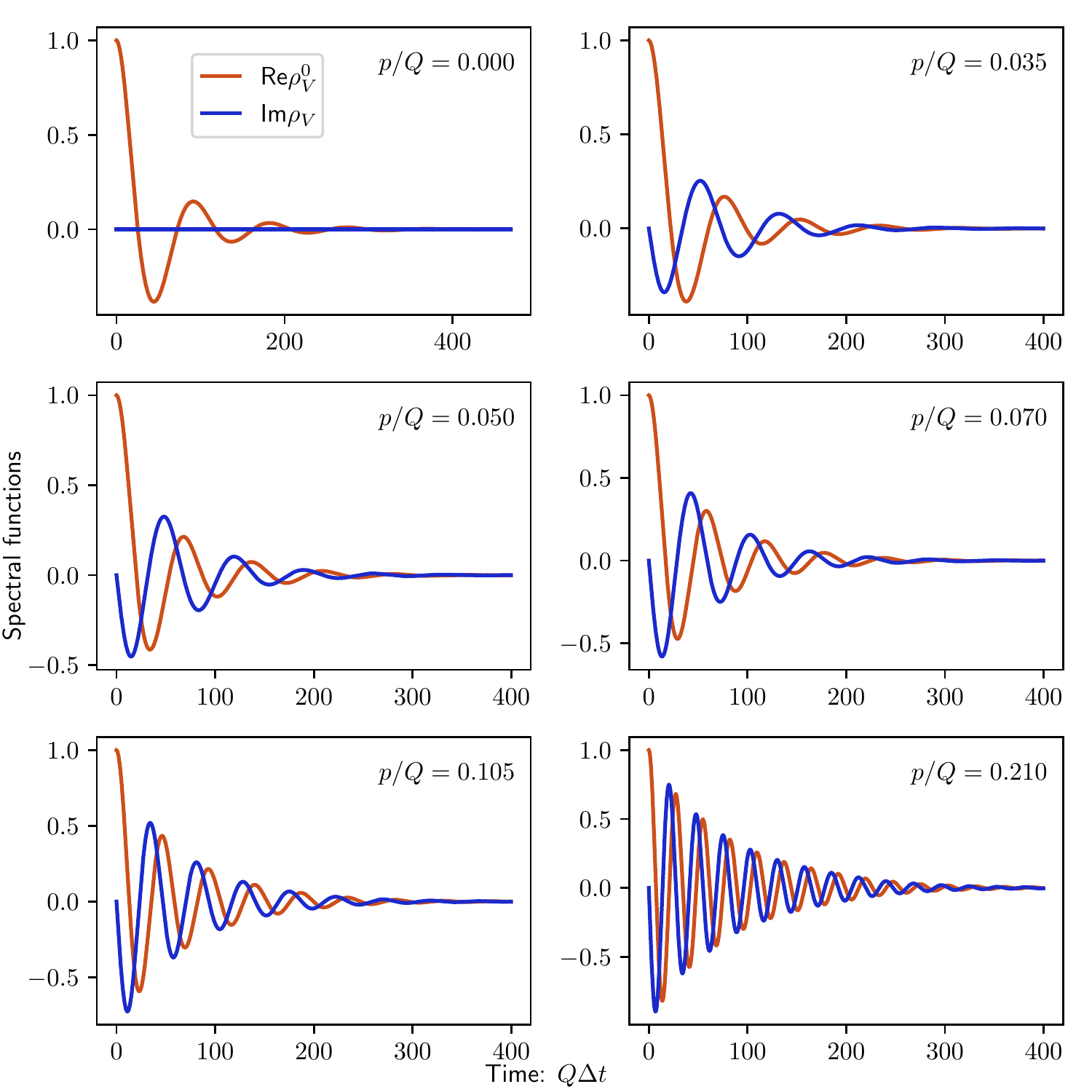}
 \includegraphics[width=0.47\textwidth]{\pToFigs/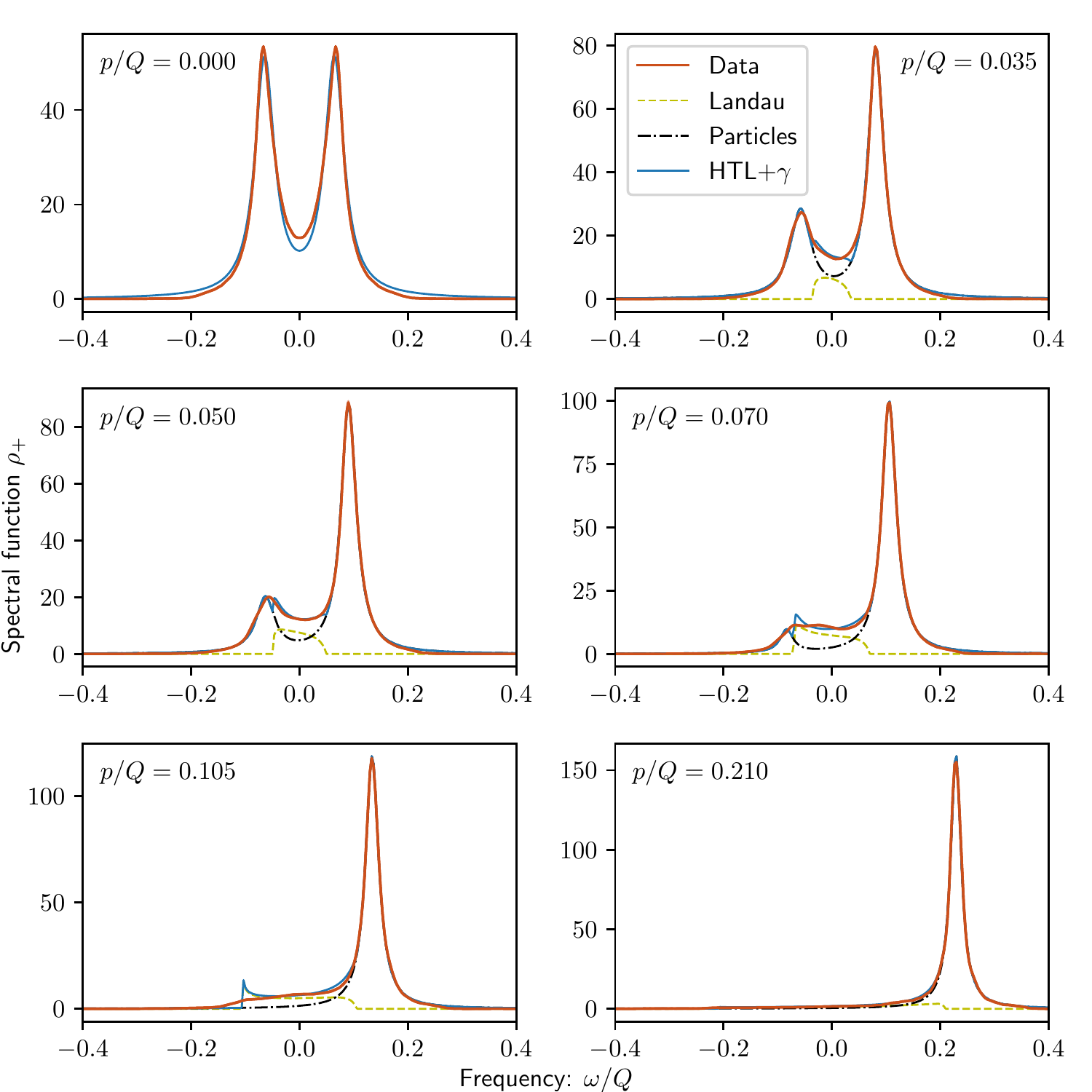}
 \caption{{\em Left:} Components of the spectral function $\rho(\tpert{+}\Delta t, \tpert, p)$ at different momenta. {\em Right:} The spectral function $\rho_+(t,\omega,p) \equiv \rho_V^0 + \rho_V$ in frequency space for the same momenta. The values at negative frequency stem from $\rho_+(t,-\omega,p) = \rho_-(t,\omega,p) \equiv \rho_V^0 - \rho_V$. 
 Figures taken from \re\cite{Boguslavski:2021kdd}.}
 \label{fig:rho_dt_w}
\end{figure}

\subsection{Fermion $\rho$ in 3+1D at an attractor}
\label{sec:fermion_rho_3D}

We run our simulations on a $256^3$ lattice with $\Q a_s = 0.75$  with almost vanishing fermion mass $m = 0.003125\,\Q$ to avoid ambiguities
of the Wilson-Dirac equation. The fermion spectral function is extracted at the reference time $\Q \tpert = 1500$, i.e., sufficiently late to make sure that we are close to the self-similar regime of \eq\eqref{eq:self-sim}. The (nonvanishing) vector components of the spectral function
\be
 \label{eq:rho_compon}
 \rho^{0}_{V}=\frac{1}{4} \text{Tr} (\rho \gamma^{0})\,, \quad
 \rho_{V}=- \frac{E_{\mbf p}\,p^{j}}{4\,p^2}\, \text{Tr} (\rho \gamma^{j})\,, 
\ee
are shown in the left panel of \fig\ref{fig:rho_dt_w} as functions of $\Delta t \equiv t'-t$ for different momenta. 
One observes that they exhibit damped oscillations with a well defined frequency (dispersion relation) and damping rate. 

This can be made more quantitative after Fourier transforming it with respect to $\Delta t$ into the frequency domain, as shown in the right panel of \fig\ref{fig:rho_dt_w}. One finds two quasiparticle peaks at $\omega = \pm\omega_\pm(p)$ 
with respective width (damping rates) $\gamma_\pm(p)$. To extract these values, we recall that the spectral function can be computed perturbatively in HTL as \cite{Blaizot:2001nr}
\begin{align} 
 &\rho_{+}^{\rm HTL}(\omega, p)= 2\pi\,\beta_+(\omega/p,p) \non
 & \quad +2\pi \left[ Z_+(p)\delta(\omega - \omega_{+}(p)) + Z_-(p)\delta(\omega + \omega_{-}(p)) \right],
\end{align}
with the Landau damping term $\beta_+$. We extend this expression to allow for a finite peak width, denoting it HTL+$\gamma$,
\begin{align} 
 \label{eq:HTL+}
 &\rho_{+}^{{\rm HTL} + \gamma}(\omega, p)= 2\pi\,\beta_+(\omega/p,p) \\
 & \quad +\frac{2Z_{+}(p) \gamma_{+}(p)}{(\omega-\omega_{+}(p))^2+\gamma_{+}^2(p)} +\frac{2Z_{-}(p) \gamma_{-}(p)}{(\omega+\omega_{-}(p))^2+\gamma_{-}^2(p)}, \nonumber 
\end{align}
and show it in the right panel of \fig\ref{fig:rho_dt_w}, together with its Landau damping and quasiparticle contributions separately. One finds very good agreement with our data. The good agreement with our data thus allows us to extract $\omega_{\pm}(t,p)$, $Z_{\pm}(t,p)$ and $\gamma_{\pm}(t,p)$ by fitting to \eqref{eq:HTL+}. 

\begin{figure}[t!]
 \centering
 \includegraphics[width=0.32\textwidth]{\pToFigs/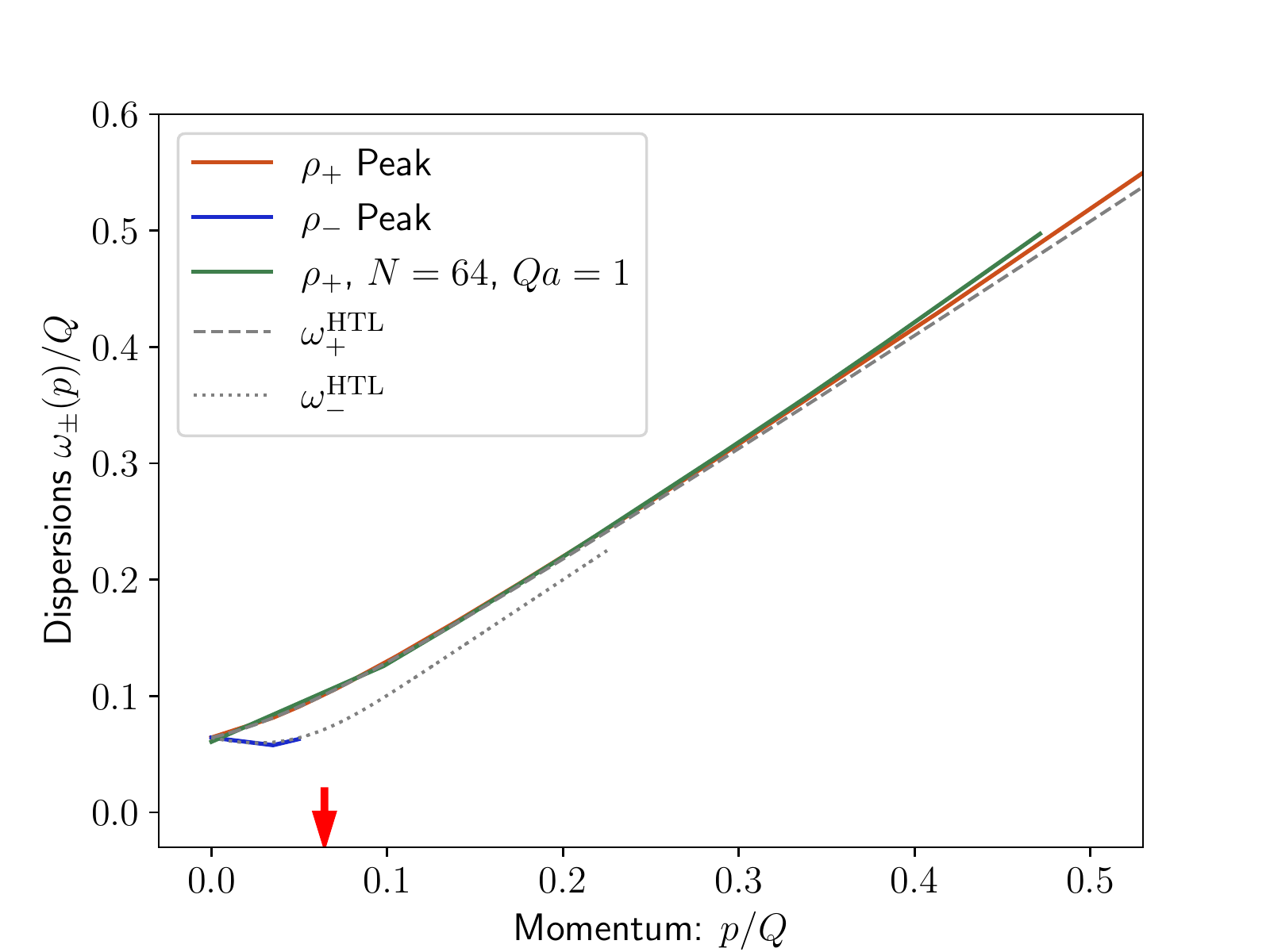}
 \includegraphics[width=0.32\textwidth]{\pToFigs/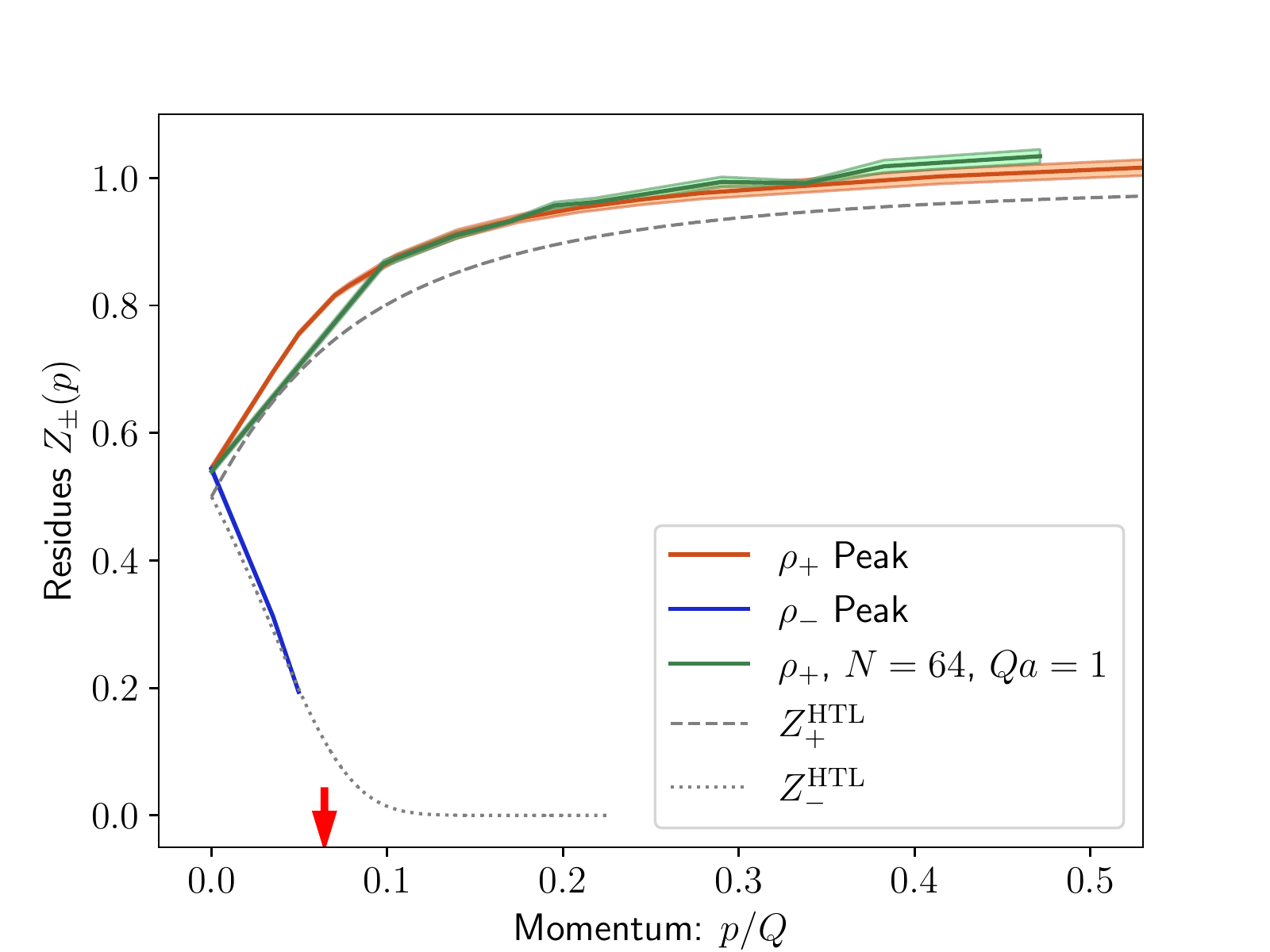}
 \includegraphics[width=0.32\textwidth]{\pToFigs/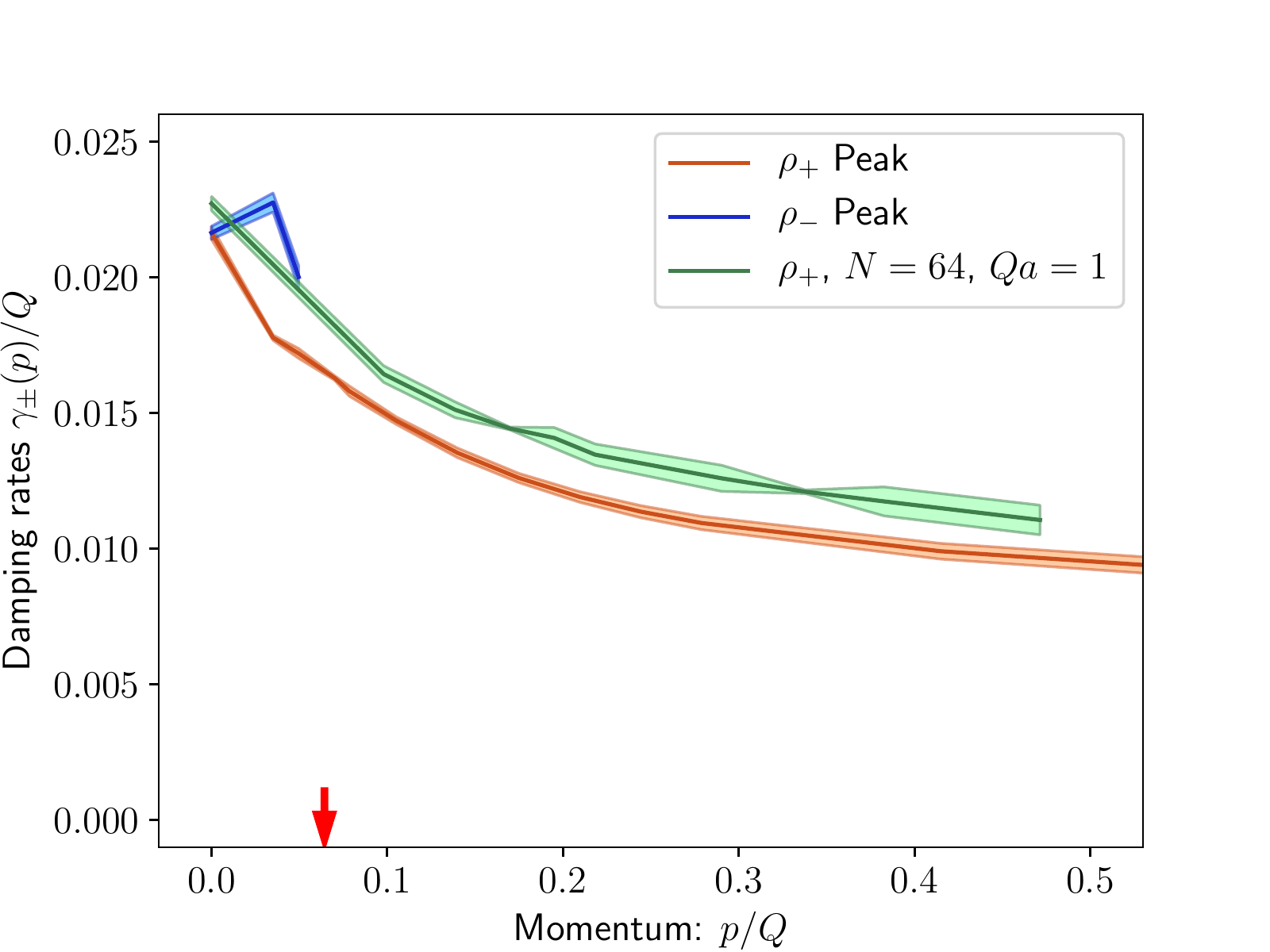}
 \caption{Extracted dispersion relations $\omega_{\pm}(t,p)$, residues $Z_{\pm}(t,p)$ and damping rates $\gamma_{\pm}(t,p)$ as functions of $p$ by fitting to \eqref{eq:HTL+}. Figures taken from \re\cite{Boguslavski:2021kdd}.}
 \label{fig:wZg_fits}
\end{figure}

These extracted quantities are shown in \fig\ref{fig:wZg_fits}, together with results from a smaller and coarser $64^3$ lattice with $Q a_s = 1$ for comparison. The HTL predictions for the dispersion relations $\omega_{\pm}^\HTL(p)$ and residues $Z_{\pm}^\HTL(p)$ are included as dashed lines. Note that they are completely determined after calculating the HTL fermion mass 
\be
 m_F = \left[C_F \int \dfrac{\ud^3 p}{(2\pi)^3}\, \dfrac{g^2f_g(p)}{p}\right]^{1/2} 
\ee
whose value is indicated by red arrows. Hence, the nice agreement between our simulation results and leading order HTL expressions shows that these quantities are dominated by perturbative physics.

In addition, our simulations allow for a first-principles insight into the momentum dependence of damping rates $\gamma_{\pm}(p)$, which is hard within a perturbative treatment. In general, we find that $\gamma_{+}(p)$ decrease with momentum and that its values are much smaller than the respective dispersion $\omega_+(p)$, validating a quasiparticle picture for (almost) all momentum modes. 

\begin{figure}[t]
 \centering
 \includegraphics[width=0.47\textwidth]{\pToFigs/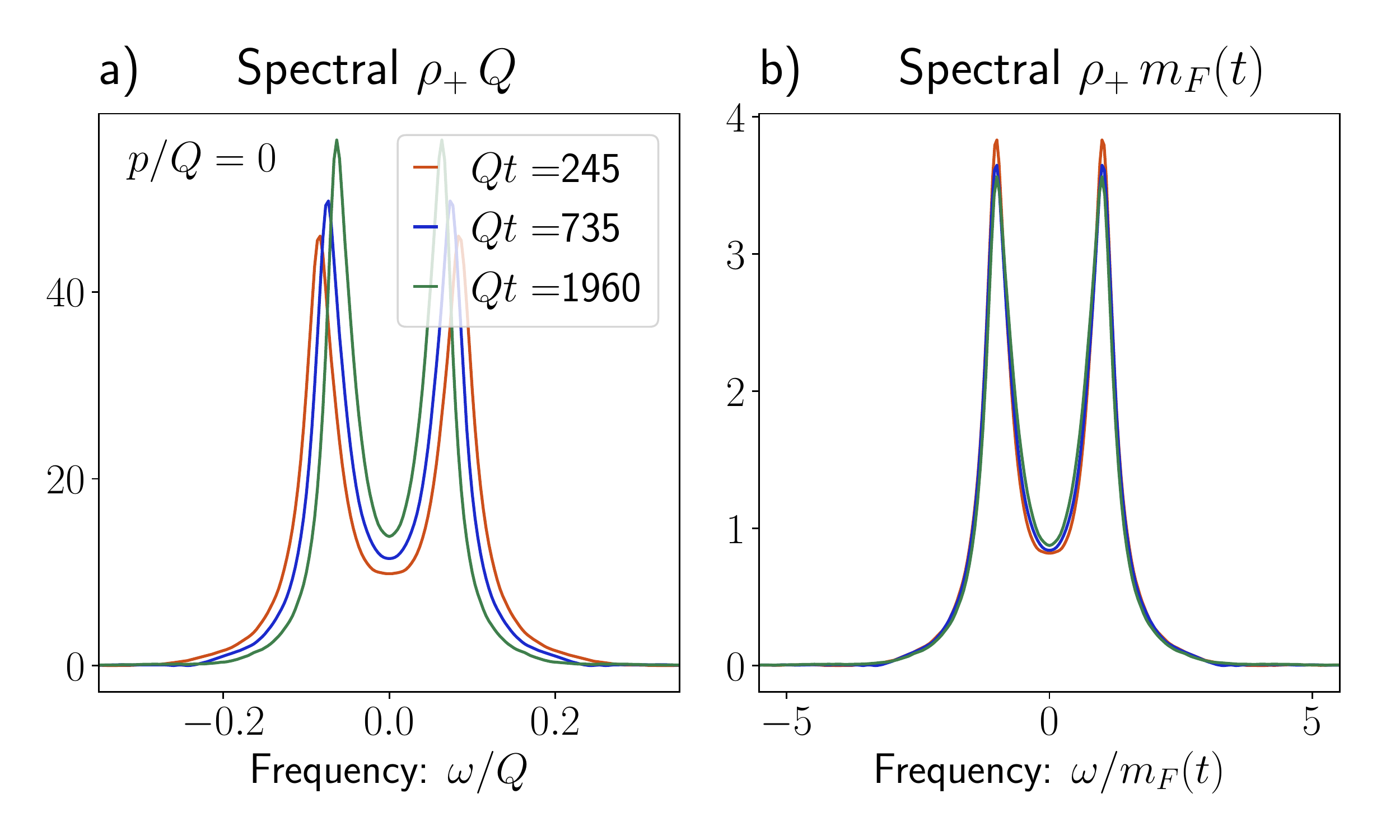}
 \includegraphics[width=0.47\textwidth]{\pToFigs/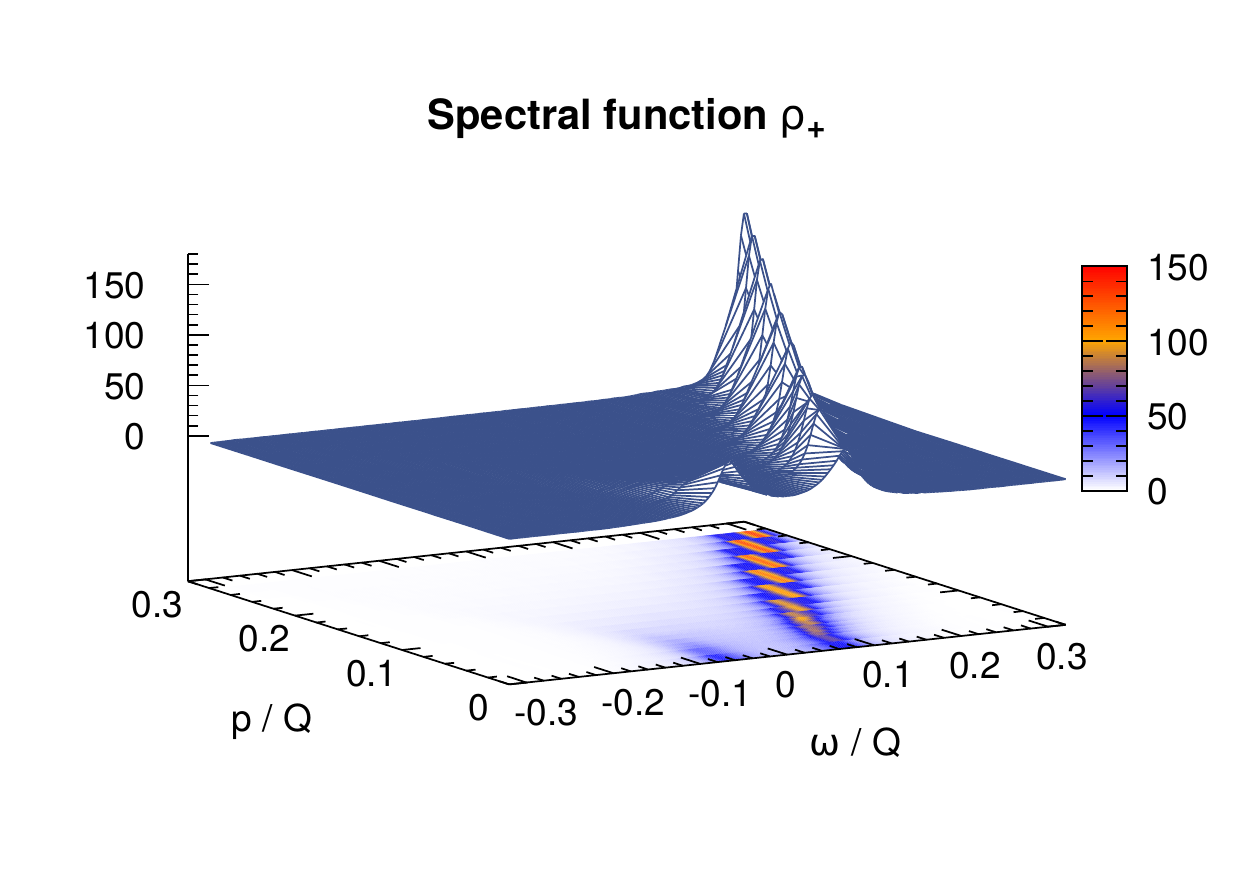}
 \caption{{\em Left: }Fermion spectral function $\rho_+(t, \omega, p{=}0)$ at different times in terms of {\em a)} the constant scale $Q$ and {\em b)} the time-dependent fermion mass $m_F(t)$.
 {\em Right:} A summary of our results: $\rho_+$ as a function of frequency and momentum.
 Figure taken from \re\cite{Boguslavski:2021kdd}.}
 \label{fig:rho_selfsim}
\end{figure}

From HTL \cite{Braaten:1992gd} one would expect the damping rate to scale with the effective temperature $\gamma^\HTL(t,p{=}0) \propto g^2 T_*(t) \sim \Q (\Q t)^{-3/7}$. Instead, we find that it rather scales with the fermion mass $\gamma(t,p{=}0) \sim m_F(t) \sim \Q (\Q t)^{-1/7}$, as we show in the left panel of \fig\ref{fig:rho_selfsim}. By rescaling all dimensionful quantities with $m_F(t) \equiv \omega_\pm(t,p{=}0)$, all curves corresponding to different (reference) times $\tpert$ collapse on a single curve. Note that this also confirms a self-similar evolution of the spectral function at the attractor. 

\section{Gluonic spectral functions}
\label{sec:gluon_rhos}


We will now discuss the gluon spectral function $\rho (x', x) = i\left\langle \left[ \hat{A}(x'), \hat{A}(x) \right] \right\rangle$ in different dimensions at corresponding self-similar attractors. Its measurement follows a very similar algorithm to the one for fermions and combines classical-statistical simulations with linear response theory. We refer to \res\cite{Boguslavski:2018beu,Kurkela:2016mhu,Boguslavski:2021buh} for more details. 
Closely related algorithms were also applied to scalar theories at self-similar attractors in \cite{PineiroOrioli:2018hst,Boguslavski:2019ecc}. 

We will discuss the color averaged transversely polarized (dotted) gluonic spectral functions $\dot\rho_T(t',t) = \partial_{t'} \rho_T(t',t)$. 
One obtains $\dot \rho_T(t,\omega)\approx \omega \rho_T$ by Fourier transforming in $\Delta t \equiv t'-t$. 

\begin{figure*}[t]
\centering
\includegraphics[width=0.45\textwidth]{\pToFigs/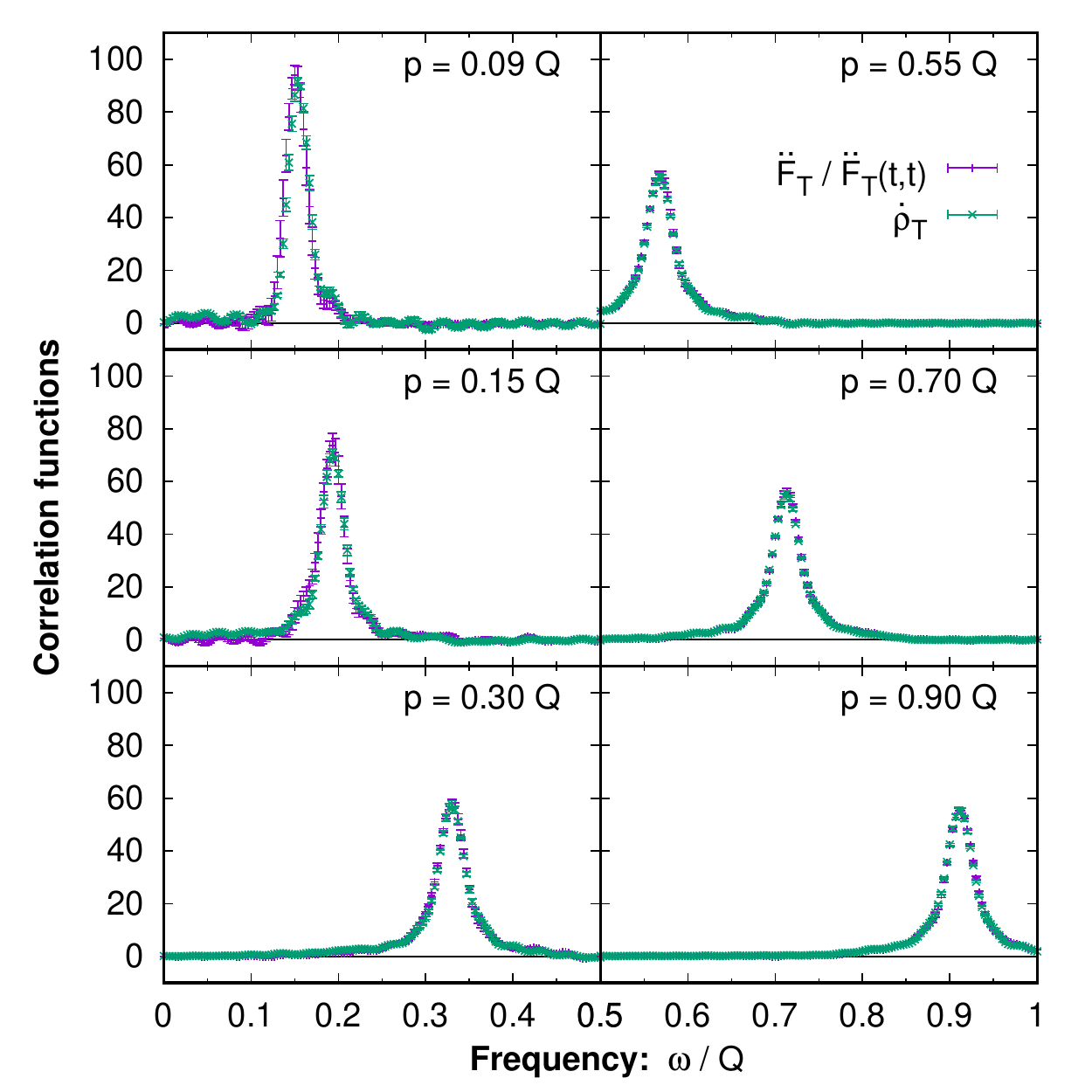}
\qquad
\includegraphics[width=0.48\textwidth]{\pToFigs/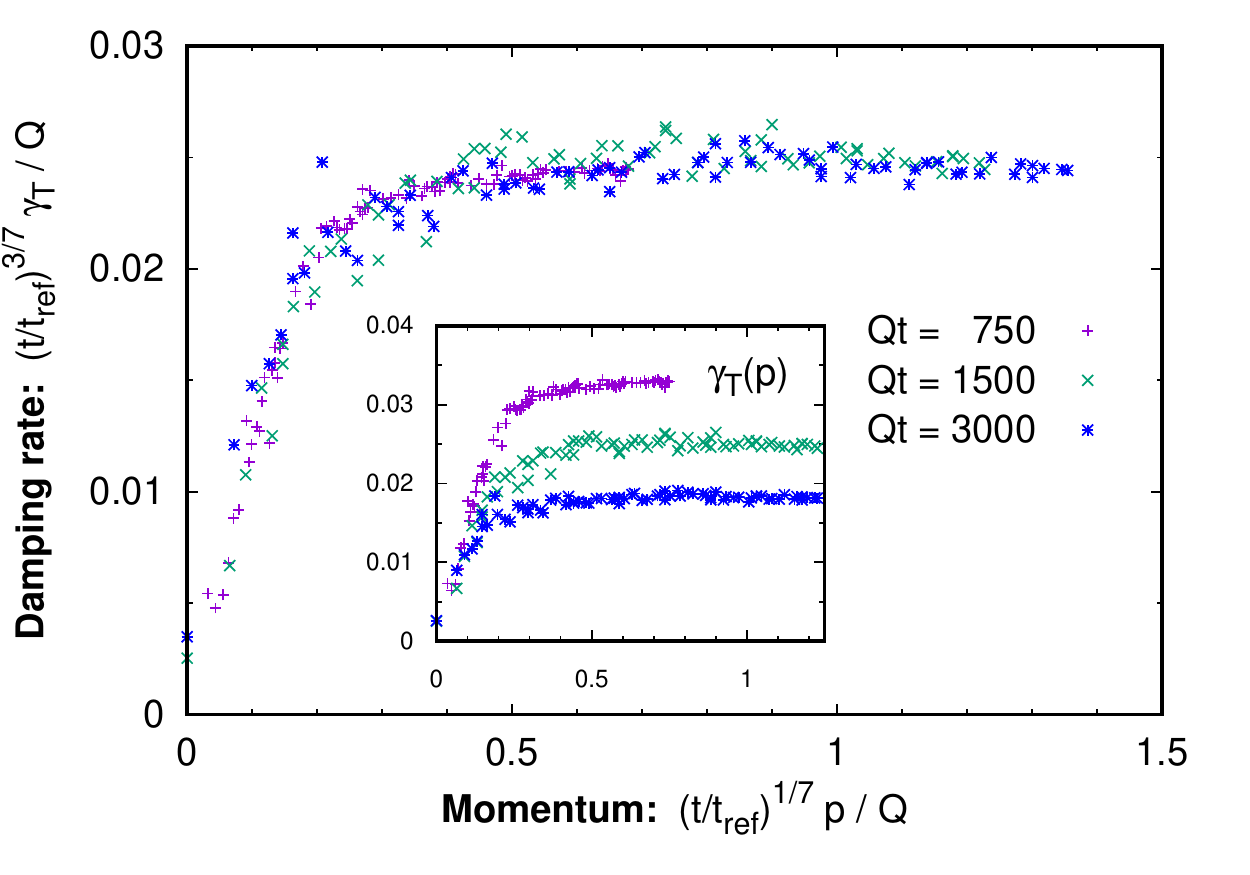}
\caption{{\bf Gluonic} simulation results in a {\bf 3+1D} plasma. 
{\em Left:} The spectral function at different momentum modes as a function of $\omega$. In contrast to the right panel of \fig\ref{fig:rho_dt_w}, we show only the positive frequency part for the gluon spectral function, which agrees with the values at negative frequency due to $\dot\rho_T(t,-\omega,p) = \dot\rho_T(t,\omega,p)$.
{\em Right:} The extracted damping rate at different times as a function of momentum, rescaled by powers of time (main plot) and without rescaling (inset). Figures taken from \re\cite{Boguslavski:2018beu}.
}
\label{fig:rho_3D}       
\end{figure*}

\begin{figure*}[t]
\centering
\includegraphics[width=0.45\textwidth]{\pToFigs/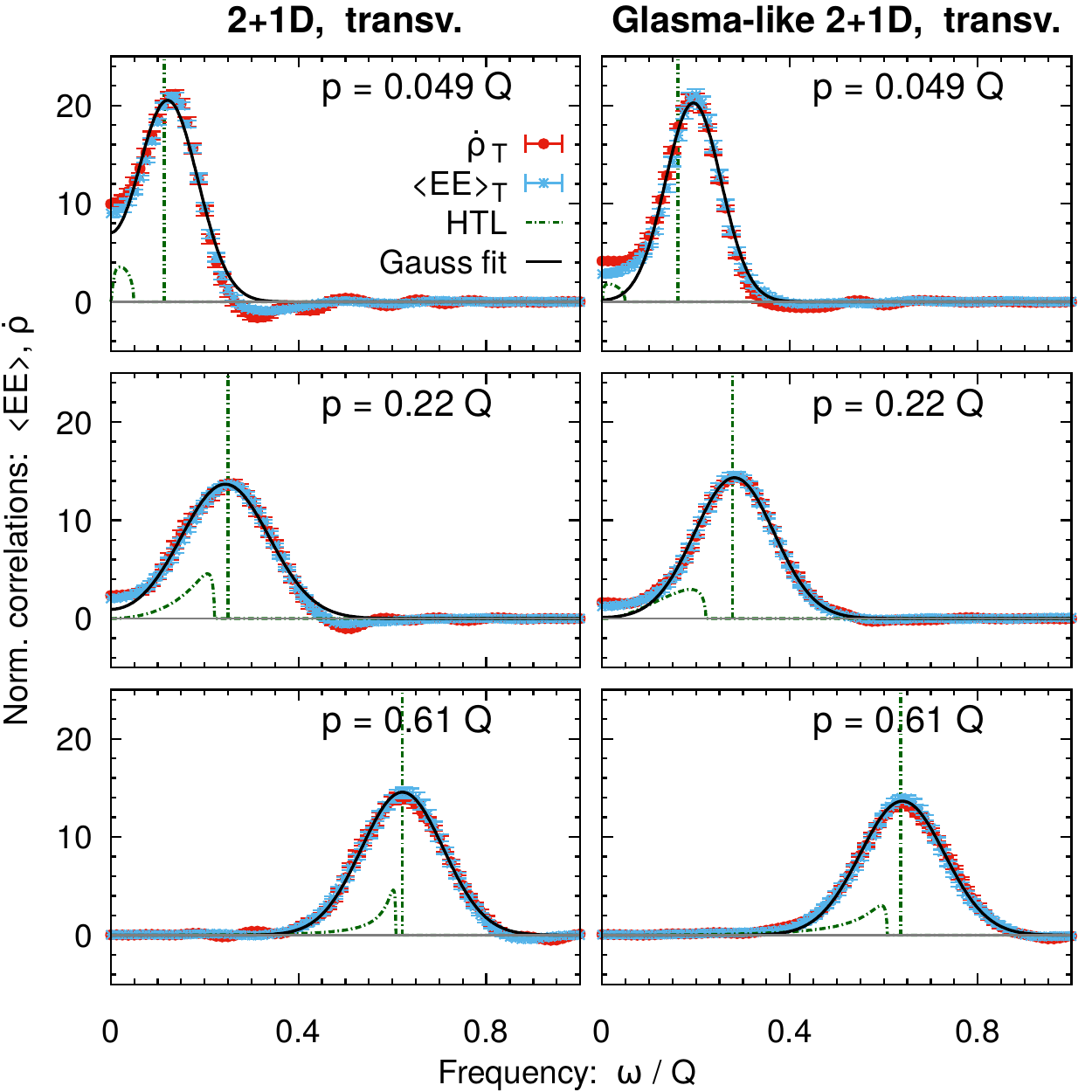}
\qquad
\includegraphics[width=0.48\textwidth]{\pToFigs/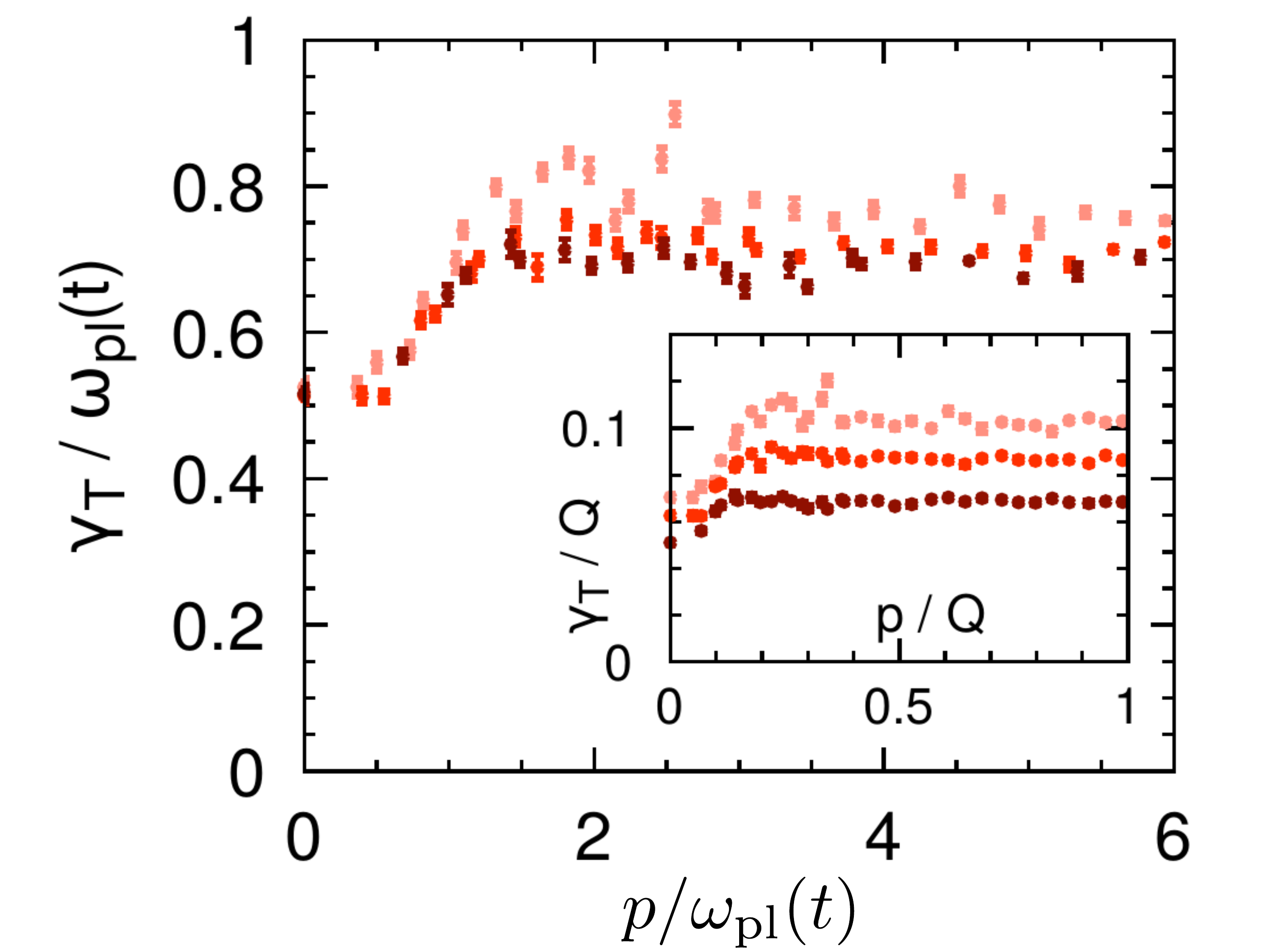}
\caption{{\bf Gluonic} simulation results in {\bf 2+1D} plasmas.
{\em Left:} The transversely polarized spectral function at different momentum modes as a function of $\omega$ for a 2+1D (left) and a 2+1D simulation coupled to an adjoint scalar field (right). 
{\em Right:} The extracted damping rate of the genuinely 2+1D plasma at different times as a function of momentum, rescaled by the time-dependent plasmon mass $\wplas(t)$ (main plot) and without rescaling (inset). Figures taken from \re\cite{Boguslavski:2021buh}.
}
\label{fig:rho_2D}       
\end{figure*}

Apart from the spectral function 
one can also measure the statistical correlation function $\langle E(t') E(t) \rangle \equiv \ddot{F}(t',t)$. While in thermal equilibrium it is related to the spectral function via the fluctuation-dissipation relation (FDR), far from equilibrium it was found numerically in \res\cite{Boguslavski:2018beu,Boguslavski:2021buh} that statistical and spectral correlation functions are related via a generalized FDR (also seen in the left panels of \figs\ref{fig:rho_3D} and \ref{fig:rho_2D}).

\subsection{Gluon spectral function in isotropic 3+1D plasmas}
\label{sec:gluon_rho_3D}

For the 3+1D simulations, the same lattice parameters and initial condition are used as in \se\ref{sec:fermion_rho_3D} (see \re\cite{Boguslavski:2018beu} for details on the gluonic simulation results).
The gluonic spectral function $\dot\rho_T \approx \omega \rho_T$ is shown in the left panel of \fig\ref{fig:rho_3D}, extracted at the same reference time $\Q \tpert = \Q t_{\text{ref}} = 1500$ as we did for the fermion $\rho_+$. One finds that $\dot\rho_T$ is dominated by narrow Lorentzian quasiparticle peaks. The damping rate (peak width) is hence seen to be much smaller than the dispersion (peak position) $\gamma_T(t,p) \ll \omega_T(t,p)$. It is further shown in \re\cite{Boguslavski:2018beu} that HTL provides an accurate description, with the dispersion relation $\omega_T(p)$, the quasiparticle residue $Z_T(p)$ and the Landau damping contribution for $|\omega| \leq p$ (here suppressed by one power of $\omega$) agreeing well with HTL predictions. 

Similarly to the fermionic case in \eq\eqref{eq:HTL+}, the damping rate $\gamma_T(t,p)$ is extracted from the gluon spectral function by fitting Lorentzian shapes to the data. The results are shown in the right panel of \fig\ref{fig:rho_3D}. The full momentum dependence is obtained and plotted at different times in the inset. Rescaling it as $\gamma_T(t,p)/g^2 T_*$, with the effective temperature $g^2 T_* \sim Q (t/t_{\text{ref}})^{-3/7}$, and plotting it as a function of $p/\wplas(t)$, with the plasmon mass $\wplas(t) \equiv \omega_T(t,p{=}0) \sim m_F(t) \sim Q (t/t_{\text{ref}})^{-1/7}$, all the curves are seen to fall on top of each other. Hence, as time grows the ratio $\frac{\gamma_{T}(t,p)}{\omega(t,p{=}0)} \sim (Q t)^{-2/7}$ decreases. This shows that the peaks become even more narrow with time and quasiparticle approximations remain valid. 


\subsection{Gluon spectral function in 2+1D plasmas}
\label{sec:gluon_rho_2D}

In \se\ref{sec:gluon_rho_3D} we considered a 3+1D plasma with a gluonic distribution $f(t,p)$ that is isotropic in momentum space. If we now allow for momentum anisotropy with the extreme case being $f(t,p_T, p_z{=}0)$, the system becomes effectively 2+1 dimensional. Here we can distinguish between a genuinely 2+1D plasma consisting of only gluonic excitations and one that is coupled to an adjoint scalar field, which we refer to as Glasma-like. The latter corresponds to the case of extreme momentum anisotropy mentioned above.

We consider these 2+1D systems far from equilibrium and discuss the spectral functions close to their self-similar attractors (see \re\cite{Boguslavski:2021buh} for details). The numerical results are summarized in \fig\ref{fig:rho_2D}. In the left panel $\dot\rho_T$ is shown in the frequency domain for different momenta. Different from gluonic and fermionic spectral functions in 3+1D, it exhibits broad non-Lorentzian peaks of approximately Gaussian shape and differs considerably from perturbative HTL predictions that are displayed as green dashed lines. 
The failure of HTL perturbation theory is not surprising in 2+1D. There the HTL approximation fails because soft momentum modes provide important contributions 
\cite{Boguslavski:2019fsb}.

The damping rate $\gamma_T(t,p)$ is extracted by performing Gaussian fits to the data and is shown at different times in the right panel of \fig\ref{fig:rho_2D}. Rescaled by the time-dependent plasmon mass $\wplas(t)$, the damping rate becomes stationary, showing $\gamma(t,p) \sim \wplas(t)$ even quantitatively. Hence, the damping rate remains of the order of $\wplas(t)$. Since the dispersion is $\omega_T(t,p) \approx \wplas(t)$ for modes $p \lesssim \wplas(t)$, quasiparticle approximations break down for these soft momenta. 


\section{Comparison and conclusion}
\label{sec:conclusion}

We have presented a new tool to extract the fermion spectral function in highly occupied plasmas nonperturbatively and we have revisited a similar method for gluonic spectral functions. They enable a first-principles determination of damping rates and were applied to classical self-similar attractors in different dimensions. 

The extracted fermion and gluon spectral functions share many properties in 3+1D. They exhibit Lorentzian quasiparticle peaks and are well described by perturbative HTL expressions. Their damping rates are in general much smaller than the dispersions, confirming a valid quasiparticle picture. However, there are also interesting differences in the momentum dependence of the damping rates. While the gluonic damping rate $\gamma_{T}$ increases with momentum, the fermionic one $\gamma_{+}$ decreases. Moreover, the gluonic damping rate decreases in time faster than the mass, while the fermionic one scales with the mass for $p=0$. 

Interestingly, the scaling $\gamma(t,p{=}0) \sim \wplas(t)$ has also been observed for the gluonic spectral function in 2+1D systems (see \se\ref{sec:gluon_rho_2D}). 
However, there the peaks in the spectral function are generally much broader than in 3+1D and a perturbative HTL description breaks down. 

So far, we have applied the methods to systems in Minkowski space-time. In the future, we want to apply our framework to Bjorken expanding (anisotropic) systems and to heavy quarks, which are of phenomenological relevance to heavy-ion collisions. Moreover, nonperturbative properties of the spectral or statistical correlators out of equilibrium can have a sizable impact on transport coefficients, as was shown in \cite{Boguslavski:2020tqz}. Thus, it will be interesting to study how observables like the photon production rate are affected by our numerical results.

\bibliography{spires}
%

\end{document}